\documentclass[aps,prm,reprint,superscriptaddress,amsmath]{revtex4-2}
\usepackage{lipsum}
\usepackage{graphics} 
\usepackage{epsfig}
\usepackage{subfigure}
\usepackage{textcomp}
\usepackage{setspace}
\usepackage{amsmath}

\begin{document}
\bibliographystyle{apsrev4-1} % Choose Phys. Rev. style for bibliography, Rev.4
%Title of paper
\title{Equation of state of rhenium under high temperatures and pressures predicted by ensemble theory}
\author{Yue-Yue Tian}
\affiliation{Institute of Modern Physics, Fudan University, Shanghai 200433, China}
\affiliation{Applied Ion Beam Physics Laboratory, Fudan University, Shanghai 200433, China}
\author{Hui-fen Zhang}
\affiliation{Key Laboratory of Thorium Energy, Chinese Academy of Sciences, Shanghai 201800, China}
\affiliation{Shanghai Institute of Applied Physics, Chinese Academy of Sciences, Shanghai 201800, China}
\author{Bo-Yuan Ning}
\email{ningboyuan@sinap.ac.cn}
\affiliation{Key Laboratory of Thorium Energy, Chinese Academy of Sciences, Shanghai 201800, China}
\affiliation{Shanghai Institute of Applied Physics, Chinese Academy of Sciences, Shanghai 201800, China}
\author{Xi-Jing Ning}
\email{xjning@fudan.edu.cn}
\affiliation{Institute of Modern Physics, Fudan University, Shanghai 200433, China}
\affiliation{Applied Ion Beam Physics Laboratory, Fudan University, Shanghai 200433, China}
\date{\today}

\begin{abstract}
The high-temperature and high-pressure equations of states (EOSs) of rhenium up to 3000 K and 900 GPa are predicted by a recently developed method in the framework of statistical ensemble theory with \textit{ab initio} computational precision. The predicted isothermal EOSs are generally consistent with semi-empirical calculations below 150 GPa and 3000 K. Especially, the predicted isobaric EOS at one atmosphere is in good agreement with previous experiments. Moreover, the bulk modulus obtained in this work is closer to the experimental measurements than other theoretical works. Based on our calculations, the disputes between previous experiments are analyzed, and it is expected that the EOSs predicted under extreme conditions might be verified in future experiments.
\end{abstract}


\maketitle
\section{Introduction}
\label{sec:intro}
The rare metal of rhenium (Re) has unique applications to high-temperature materials due to its high melting point ($\sim$ 3453 K), ductility and diffusional viscosity, for instance, to mediate the fragility of fusion nuclear materials~\cite{2018-w}, or to increase the high-temperature deformation resistance of turbine blade materials~\cite{2022-re,2009-re}. In addition, the large bulk modulus of Re makes it play a vital role in high-pressure researches. For example, Re is commonly selected to be as the gasket material in the diamond-anvil cell (DAC)~\cite{2012-du,Ono,2014-a} because of its high compressive strength to retain appreciable thickness up to several hundred gigapascal (GPa)~\cite{2012-sin}. As to internally heated DAC, Re is an excellent choice for containing and heating samples, and thus, is always utilized as an internal pressure gauge for determining pressures in high-temperature X-ray diffraction experiments~\cite{2004-zha}. Because of its specialties in the realm of extreme-condition physics researches, it is indeed necessary to acquire a reliable equation of state (EOS) which can readily determine the thermodynamic properties of Re.

Although investigations on EOSs of Re started about fifty years ago~\cite{1970-liu,1987-vohra}, there still exist controversies over the EOS under high pressures at ambient temperature among recent measurements~\cite{Ono,2021-sa,2014-a,2004-zha,2012-du,2018-sakai} and the high-temperature EOSs of Re have not been much studied extensively. Dubrovinsky \textit{et al}.~\cite{2012-du} used static double-stage DAC (ds-DAC) to compress Re and achieved a pressure record of 640 GPa at room temperature for the first time. However, the obtained EOS in Ref.\cite{2012-du} exhibits about 13\% difference between the one measured by Anzellini \textit{et al}.~\cite{2014-a}. Such an inconsistence was also noted by another ds-DAC experiment conducted later by Sakai \textit{et al}.~\cite{2018-sakai}. In a detailed analysis, Saikai \textit{et al}. pointed out that the two EOSs have large deviations and the corresponding pressures are $630$ GPa and $430$ GPa based on the two pressure calibrations in Refs.~\cite{2012-du} and~\cite{2014-a} respectively when the volume is compressed to $V/V_0=0.633$.

Theoretical tools of first-principles quantum mechanics may provide helpful insights to tackle with the problems concerning extreme-condition measurements. For the EOS of Re, predictions based on either molecular dynamics simulations~\cite{2018-bur} or static energy calculations~\cite{2003-verma,2022-xian,2022-xian} coincided with the experimental data~\cite{2014-a,2012-du,2021-sa,1970-liu,1987-vohra,2012-lv} in the lower pressure zone ($\leq 150$ GPa) but showed obvious divergence in the higher pressure zone. Rech \textit{et al.}~\cite{2019-rech,2020-rech} very recently used both full-potential all-electron and pseudopotential-based density functional theory (DFT) to conduct calculations covering a wide pressure range up to $1500$ GPa and the results were in favor of the EOS measured by Refs.~\cite{2014-a} and~\cite{1987-vohra}. However, it should be noted that most of the above computations were implemented at 0 K, which may be insufficient to include finite-temperature thermodynamic effects on the EOS, let alone the high-temperature situations. 

The statistical ensemble theory in fact has already formulated a complete theoretical framework to include the thermal effects and paved a strict way to directly derive EOS at any temperature and pressure by calculating the partition function (PF) without any parameters or empirical hypothesis~\cite{holzapfel1998,fe,pes1}. Unfortunately, the exact solution to the PF of a system consisting of \textit{N} particles involves a 3\textit{N}-fold configurational integral, the computational cost of which makes it difficult to implement the advanced first-principles computations. Multiple methods have been developed to solve the calculation problem of PF, such as the phonon model based on quasi-harmonic approximation (QHA)~\cite{fultz2010,ceder2002,2001-deb,2017-pal}, particle in a cell model (PIC)~\cite{1996-wass,2001-cohen,2003-gann}, classical mean-field (CMF) approach~\cite{2000-cmf,2000-wang,2002-wang,2005-wang,2006-wang,2001-wang} an so on. However, the accuracy of QHA was questioned for not sufficiently considering the anharmonic vibrations of atoms at high temperature~\cite{hellman2011} or with large atomic volumes~\cite{2019-gong}. The accuracy of PIC may also be limited due to the neglects of interatomic correlations and diffusions, i.e., the vibrations of atoms are independent with each other in the framework of PIC~\cite{1996-wass,2001-cohen,2003-gann}, and previous works~\cite{2024-tian,2022-han} already showed less precision of the CMF model, which is further simplification to the configurational integral on the basis of the PIC model.

Different from previous methods, we recently developed a direct integral approach (DIA) to the PF without any empirical or artificial parameters~\cite{2021-ning}, and DIA has been successfully applied to calculate the EOSs of solid copper~\cite{2021-ning}, argon~\cite{2019-gong,2021-gong}, 2-D materials~\cite{DIA-2D}, tungsten~\cite{2024-tian} and reproduce phase transition of metal vanadium~\cite{dia-v}, zirconium~\cite{dia-zn} and aluminum~\cite{dia-al}. Especially, a recent work concerning with the EOS of iridium~\cite{2022-han} showed that the results from DIA are much more precise than CMF and QHA.
In the present work, we applied DIA to solve the PF of crystal hexagonal close-packed (\textit{hcp}) Re and predicted the high-temperature and high-pressure EOSs in the range up to 3000 K and 900 GPa. We firstly compared room-temperature EOS predicted by DIA with existing literatures up to 900 GPa and settled down the disputes over the pressure calibration among experiments. Then the isothermal EOSs and bulk modulus were predicted by DIA in a range of 3000 K and 150 GPa, and compared with limited literatures. The structure of the paper is as following: In Section~\ref{sec:1}, details of the implementation of DIA and parameters of DFT are elaborated. In Section~\ref{sec:2}, careful comparisons and a discussion of the calculated results with the available experiments are presented, and conclusional remarks are finally made in the Section~\ref{sec:conclu}.

\section{Theoretical Model and Computation Method\label{sec:1}}
\subsection{Direct Integral Approach to PF}
\label{sec:1:1}
For a system containing \textit{N} atoms confined within volume \textit{V} at temperature \textit{T}, the atoms are regarded as \textit{N} point particles with cartesian coordinate $\textbf{\textit{q}}^{N}=\{\textbf{\textit{q}}_{1},\textbf{\textit{q}}_{2},\cdots \textbf{\textit{q}}_{N}\}$ and the total energy, $U(\textbf{\textit{q}}^{N})$, as the function of $\textbf{\textit{q}}^{N}$ can be computed by quantum mechanics. With the knowledge of $U(\textbf{\textit{q}}^{N})$, PF of the system reads
\begin{equation}
Z=\frac1{N!}(\frac{2\pi m}{\beta h^{2}})^{\frac32N}Q, \label{eq1}
\end{equation}
with configurational integral
\begin{equation}
Q = \int d{\textbf{\textit{q}}}^{N} e^{-\beta U({\textbf{\textit{q}}}^{N})}, \label{eq2}
\end{equation}
where \textit{h} is the Planck constant and $\beta = 1/k_BT$ with $k_B$ the Boltzmann constant. If the configurational integral \textit{Q} is solved, the pressure (\textit{P}), 
\begin{equation}
P=\frac{1}{\beta}\frac{\partial lnZ}{\partial V},\label{eq3}
\end{equation}
Helmholtz FE (\textit{F}), 
\begin{equation}
F=-\frac{1}{\beta}lnZ,\label{eq4}
\end{equation}
and all of the other thermodynamic quantities can be obtained.

For a crystal with \textit{N} atoms at their ideal lattice sites $\textbf{\textit{Q}}^N=\{\textbf{\textit{Q}}_1,\textbf{\textit{Q}}_2,\cdots \textbf{\textit{Q}}_N\}$ and the total energy $U_0(\textbf{\textit{Q}}^N)$, we firstly introduce a transformation,
\begin{equation}
\textbf{\textit{q}}^{\prime N}=\textbf{\textit{q}}^N-\textbf{\textit{Q}}^N, U^{\prime}(\textbf{\textit{q}}^{\prime N})=U(\textbf{\textit{q}}^N)-U_0(\textbf{\textit{Q}}^N),\label{eq5}
\end{equation}
where $\textbf{\textit{q}}^{\prime N}$ and $U^{\prime}(\textbf{\textit{q}}^{\prime N})$ represent the displacements of atoms away from their ideal lattice positions and the corresponding total energy differences with respect to $U_0(\textbf{\textit{Q}}^N)$. By inserting Eq. (5) into Eq. (2), configurational integral \textit{Q} is expressed as
\begin{equation}
Q=e^{-\beta U_0}\int d\textbf{\textit{q}}^{\prime N} e^{-\beta U^{\prime}(\textbf{\textit{q}}^{\prime N})}.\label{eq6}
\end{equation}
Clearly, $U^\prime(\textbf{\textit{q}}^{\prime N})$ is positive within all the integral domain and has minimum at the origin $(U'(0)=0 )$. According to DIA\cite{2021-ning}, the integral in Eq. (6) equals to an effective 3\textit{N}-dimentional volume,
\begin{equation}
Q=e^{-\beta U_0}\prod_{i=1}^N(\mathcal{L}_{i_x}\mathcal{L}_{i_y}\mathcal{L}_{i_z}),\label{eq7}
\end{equation}
with the effective length
\begin{equation}
  \mathcal{L}_{ix}=\int e^{-\beta U'(0\cdots q'_{ix}\cdots0)} dq'_{ix},\label{eq8}
\end{equation}
\begin{equation}
  \mathcal{L}_{iy}=\int e^{-\beta U^{\prime}\left(0\cdots q'_{iy}\cdots0\right)} dq'_{iy},\label{eq9}
\end{equation}
and
\begin{equation}
  \mathcal{L}_{iz}=\int e^{-\beta U^{\prime}(0\cdots q'_{iz}\cdots0)} dq'_{iz},\label{eq10}
\end{equation}
where $q'_{ix}$($q'_{iy}$ or $q'_{iz}$) is the \textit{x} (\textit{y} or \textit{z}) coordinate of the \textit{i}th atom moving away off its ideal lattice site along with the other two degrees of freedom of the atom and all the other atoms fixed. 
%----------------fig.1----------
\begin{figure}
\centering\includegraphics[width=7.5 cm]{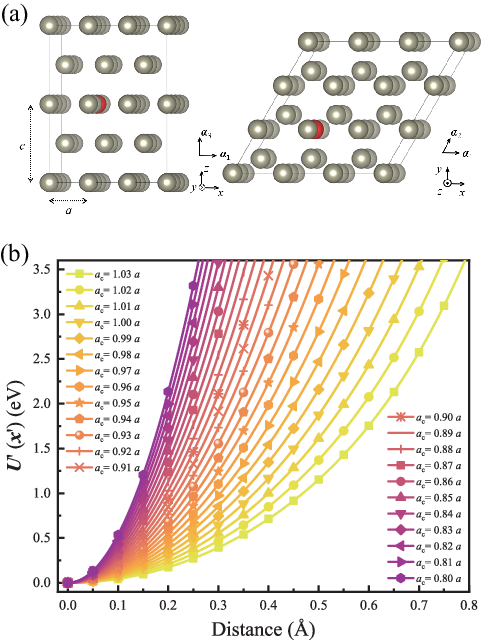}
\caption{\label{fig1} (Color Online) (a) A 3 × 3 × 2 supercell of \textit{hcp} Re, in which an arbitrary atom colored in red is moved separately along \textit{x}, \textit{y} or \textit{z} direction step by step to the final position, respectively. (b) A series of calculated $U'(x')$ (along \textit{x} direction as an example) dependent on the volume of \textit{hcp} Re structure recognized via the lattice constant $\textit{a}_{c}$ changing from 0.80\textit{a} to 1.03\textit{a}.}
\end{figure}

For \textit{hcp} Re, all the atoms are equivalent and therefore, the configurational integral is calculated as
\begin{equation}Q=e^{-\beta U_{0}}\Big(\mathcal{L}_{x}\mathcal{L}_{y}\mathcal{L}_{z}\Big)^{N}.\end{equation}
For implementing the calculations, we constructed a 3 × 3 × 2 supercell containing 36 atoms with basis vectors $\boldsymbol{\alpha_{1}}=a\cdot(1,0,0)$, $\boldsymbol{\alpha_{2}}=a\cdot(1/2, \sqrt{3}/2, 0)$ and $\boldsymbol{\alpha_{3}}=(0,0,\mathrm{\textit{c}})$, where \textit{a} and \textit{c} are the lattice parameters. The \textit{hcp} system is firstly relaxed to find the most stable structural parameters, and then, as shown in Fig. 1(a), an arbitrary atom (red one in Fig.~\ref{fig1}(a)) was chosen to move away along \textit{x} (\textit{y} or \textit{z}) direction step by step with an interval of 0.05 Å to gain the total energy $U^{\prime}(x^{\prime})$ ($U^{\prime}(y^{\prime})$ or $U^{\prime}(z^{\prime})$), which were calculated by DFT. Afterwards, the cubic spline interpolation algorithm\cite{spline} was used to smoothen the energy curve $U^{\prime}$, and the configurational integral \textit{Q} and the PF of the structures at given volume can be calculated via Eq. (8-11) and (1). Changing the volume of the structure by enlarging or shrinking the lattice constants, a series of $U^{\prime}$ as a function of volume were obtained ($U^{\prime}(x^{\prime})$ as an example is shown in Fig.~\ref{fig1}(b)). Correspondingly, a set of configurational integral \textit{Q} dependent on volume could be achieved, and further the EOSs were calculated based on Eqs. (1) and (3).

\subsection{Computational Parameters of DFT}
\label{sec:1:2}
DFT calculations in the present work were implemented in Vienna Ab initio Simulation Package~\cite{paw1-kresse,paw2-kresse}, in which Projector Augmented Wave (PAW) pseudopotential\cite{paw3-bol,paw4-kress} and Perdew Burke Ernzerhof (PBE) of general gradient approximation (GGA)\cite{gga} were selected to describe the electron-ion interactions and electron-electron exchange-correlation functional with 7 valence electrons ($6s^25d^5$) considered. A $\Gamma$-centered 11 × 11 × 11 uniform k-mesh grid is set to sample the Brillouin zone by the Monkhorst-Pack scheme together with the electron self-consistent tolerance for total energy being $1\times10^{-6}$ eV and the plane-wave energy cutoff being 400 eV. Convergence verifications of all these parameters were conducted for the structure with the smallest volume ($a_{\mathrm{c}}=0.80 a$) considered in the present work that the fluctuation of the total energy was much less than $10^{-3}$ eV/atom. In order to check the influences of different functionals on the results,
we compared the EOS calculated by the local-density approximation (LDA), PBE and PBEsol at 0 K with the experimental results and found that PBE  performs better at low pressure, as shown in Fig. S1 in the supplementary materials.

\section{Results and Discussions\label{sec:2}}
\subsection{EOS at room temperature}
\label{sec:2:1}
The EOS of Re at room temperature determined by DIA is shown as the red solid line in Fig.~\ref{fig2}(a) and the \textit{P-V} data are listed in Table S1 in the supplementary materials,
where the obtained lattice constants $a=2.780 $ {\AA} and the axial ratio $c/a=1.615$ at ambient condition are all in good agreements with previous studies~\cite{Ono,1970-liu,1987-vohra}.
The experimental EOSs measured in Refs.~\cite{Ono,2014-a,2012-du,2018-sakai} and the theoretical one at $0$ K predicted by Rech \textit{et al}.~\cite{2019-rech} using all-electron DFT are also exhibited in Fig.~\ref{fig2}(a),
and for more explicit comparisons, the pressure deviation between the experiments and DIA are shown in Fig.~\ref{fig2}(b).
As denoted in the green dashed line in Fig.~\ref{fig2}(a), the 0 K EOS by Rech \textit{et al}.~\cite{2019-rech} is similar to that by DIA and the maximum relative difference is below 5\%, which may result from the complete neglect of the finite-temperature effects in Ref.~\cite{2019-rech} and, on the other hand, validates the correctness of the parameters of PAW pseudopotential and the PBE functional used in the present work to a certain extent.
As can be seen clearly, except for the EOS from Ref.~\cite{2012-du}, the result of DIA $(P_{\mathrm{DIA}})$ is in a good agreement with the experiments $(P_{\mathrm{Exp}})$ and the relative deviation $(\mathrm{RD}=(P_{\mathrm{Exp}}-P_{\mathrm{DIA}})/P_{\mathrm{Exp}})$ between  $P_{\mathrm{DIA}}$ and other $P_{\mathrm{Exp}}$ is less than $10\%$ within 200 GPa.
%----------------fig.2----------
\begin{figure*}
\centering\includegraphics[width=15.5 cm]{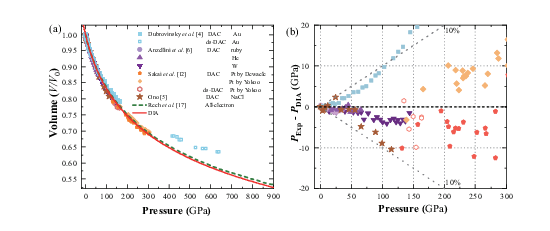}
\caption{\label{fig2}(Color Online) (a) Room temperature (300 K) \textit{P-V} results from DIA and experimental measurements. Blue solid and open squares are the results measured by Dubrovinsky \textit{et al}.\cite{2012-du}; purple circles, up and down triangles are the results measured by Anzellini \textit{et al}.\cite{2014-a}, orange pentagons, diamonds, and open hexagons are the results measured by Sakai \textit{et al}.\cite{2018-sakai}, brown stars are the results measured by Ono\cite{Ono} with $V_{0}$ being the volume at one atmosphere. The red solid and green dashed lines are the calculated results from DIA and Rech \textit{et al}.\cite{2019-rech}. (b) is the deviation of the calculated pressure by DIA from the experimental one under same $V/V_{0}$, and gray dashed lines labeled by 10\% present the relative deviation of 10\%.}
\end{figure*}

The experiment by Dubrovinsky \textit{et al}.~\cite{2012-du} consists of two parts that a conventional DAC was firstly used for compression up to 164.8 GPa and a ds-DAC was then used to accomplish a much higher pressure range of 416-640 GPa.
In the low-pressure range, the differences between $P_{\mathrm{DIA}}$ and $P_{\mathrm{Exp}}$ are about 1 GPa below 30 GPa and get to be larger with the increases of the compression that the largest relative deviation reaches 13.2\% (see blue solid squares in Fig.~\ref{fig2}(b)).
In the high-pressure range, the experimental results are even much larger than the calculated pressures that the $P_{\mathrm{Exp}}$  at the highest compression of 640 GPa is larger than $P_{\mathrm{DIA}}$ by 203.4 GPa and the corresponding relative deviation is 31.8\%.

\begin{figure*}
\centering\includegraphics[width=15.5 cm]{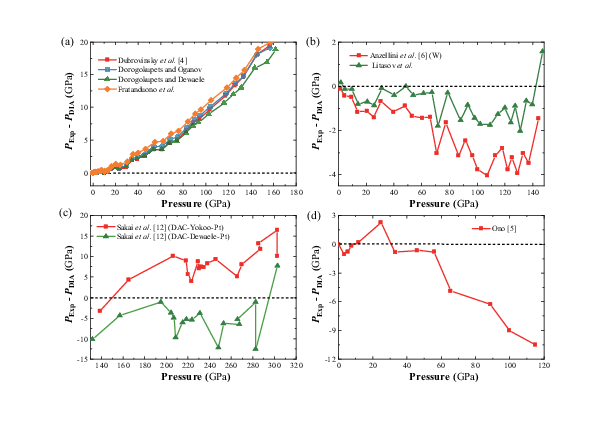}
\caption{\label{fig3}(Color Online) The deviations between the pressures determined by DIA and those by the experiments of (a) Dubrovinsky \textit{et al}.~\cite{2012-du}, (b) Anzellini \textit{et al}.~\cite{2014-a}, (c) Sakai \textit{et al}.~\cite{2018-sakai} and (d) Ono~\cite{Ono}, respectively. All the raw experimental data are denoted in red solid squares in (a)-(d). The recalibrated pressures of Ref.~\cite{2012-du} by using the Au scales proposed by Fratanduono \textit{et al}.~\cite{2021-science} and Dorogokupets \textit{et al}.~\cite{2007-dd,2007-do} are respectively displayed in orange diamonds, blue circles and green triangles in (a), the recalibrated results of Ref.~\cite{2014-a} by using the W gauge proposed by Litasov \textit{et al}.~\cite{2013-litasov} are displayed in green triangles in (b) and the results of Ref.~\cite{2018-sakai} by using the Pt gauge proposed by Dewaele~\cite{2004-dewaele} are displayed in green triangles in (c).}
\end{figure*}

In fact, the EOS determined by Dubrovinsky \textit{et al}.~\cite{2012-du} were questioned that the pressures may be grossly over estimated.
For example, Anzellini \textit{et al}.~\cite{2014-a} showed that the pressures in Ref.~\cite{2012-du} are much higher by up to 13\% when at same volumes.
It was further argued that if the EOS established by Dubrovinsky \textit{et al}. is used to calibrate the pressures, it would be overestimated by 40 GPa around 200 GPa and such an overestimation would continuously increase with the increases of pressure~\cite{2014-a}.
Sakai \textit{et al}.~\cite{2018-sakai} later reported that the compression rate ($V/V_{0}$) at 0.633 would correspond to 430 GPa based on the Re-EOS proposed by Vohra \textit{et al}.~\cite{1987-vohra} or Anzellini \textit{et al}.~\cite{2014-a}, while it would be 630 GPa according to the EOS reported by Ref. \cite{2012-du}. 
One possible reason for the obviously larger pressures in Ref.~\cite{2012-du} may be attributed to the adopted gold pressure gauge promoted by Yokoo \textit{et al}. in 2009~\cite{2009-yokoo}.
To exam this conjecture, we re-estimated the pressures of Ref.~\cite{2012-du} by other three gold scales put forward by Fratanduono \textit{et al}.~\cite{2021-science} and Dorogokupets \textit{et al}.~\cite{2007-dd,2007-do}, the latter of which are correspondingly self-consistent with the ruby and NaCl scale used respectively in the experiments by Anzellini \textit{et al}.~\cite{2014-a} (Fig.~\ref{fig3}(b)) and Ono \textit{et al}.~\cite{Ono} (Fig.~\ref{fig3}(d)).
As shown in Fig.~\ref{fig3}(a), there are no prominent changes of the differences between the re-calibrated pressures and $P_{\mathrm{DIA}}$, and thus, the uncertainty of the choices of a pressure calibrant contributes little to the obvious deviations in the experiment of Ref.~\cite{2012-du}.
Regarding to the drastically higher pressures achieved in Ref.~\cite{2012-du}, more experimental and theoretical exploring may be required to further verify the very causes in this experiment.
%----------------fig.3----------

Anzellini \textit{et al}.~\cite{2014-a} measured the EOS of Re up to 144 GPa using a conventional DAC and ran the experiments three times with helium (He)~\cite{1993-loub}, ruby~\cite{2007-do} and tungsten (W)~\cite{2007-do} as the pressure scales respectively.
Since the used ruby and tungsten gauges are self-consistent and the determined pressure range under the W scale is the largest (1.42-144 GPa), specific comparisons are performed just between the calibrated pressures by W and the calculated results.
As shown in Fig.~\ref{fig3}(b), the  $P_{\mathrm{Exp}}$ are smaller than $P_{\mathrm{DIA}}$ in the whole pressure range with the largest pressure deviation about 4 GPa.
Such a difference can be understood because the pressure scale~\cite{2007-do} used in Ref.~\cite{2014-a} was confirmed to underestimate pressures~\cite{2003-hol,2003-kunc,2004-dewaele,1978-mao}.
In 2013, Litasov \textit{et al.} established a new W-EOS~\cite{2013-litasov}, which we used to re-estimate the pressures in Ref.~\cite{2014-a} and as shown in Fig.~\ref{fig3}(b) with green triangles, the relative deviation between the redefined pressures and $P_{\mathrm{DIA}}$ is reduced to 2 GPa, which is within the experimental uncertainty. 

Sakai \textit{et al}.~\cite{2018-sakai} performed static experiments of the room-temperature EOS in a high-pressure range (100 $\sim$ 300 GPa) using Pt as the calibrant~\cite{2009-yokoo}, which is highly consistent with the Au pressure scale used in the experiment of Dubrovinsky \textit{et al}.~\cite{2012-du} while the final results are very different from that of Dubrovinsky \textit{et al}.
The pressure differences between the $P_{\mathrm{Exp}}$ of Ref. ~\cite{2018-sakai} and $P_{\mathrm{DIA}}$ are shown in Fig.~\ref{fig3}(c) as denoted in the red solid squares, with an average deviation $(\mid P_{\mathrm{DIA}}-P_{\mathrm{Exp}}\mid)$ of 8.3 GPa in the whole pressure range.
Sakai \textit{et al}.~\cite{2018-sakai} also determined the pressures with another Pt gauge promoted in 2004 by Dewaele \textit{et al}.~\cite{2004-dewaele} and the corresponding pressure differences are shown in Fig.~\ref{fig3}(c) as the green solid triangles, which shows that the experimental pressures under this scale are about 6.2 GPa smaller than $P_{\mathrm{DIA}}$ on average.
Yet, it should be pointed out here the the Pt scale established by Dewaele \textit{et al}. ~\cite{2004-dewaele} is just believed to be effective upon $\sim$ 100 GPa, and therefore, using it to calibrate pressures in a range of 100 $\sim$ 300 GPa is actually unreliable. 

In 2020, Ono~\cite{Ono} measured the EOS of Re in the pressure range of 0-115 GPa using a laser-annealing DAC and the synchrotron X-ray diffraction method.
As shown in the Fig.~\ref{fig3}(d), below 60 GPa, the pressure calculated by DIA is essentially in agreement with the measurement where the average deviation is about 0.7 GPa.
With the pressure increasing, the deviation gradually increases and reaches about 10.5 GPa at 115 GPa.
The pressure transmitting medium used in the experiment may account for the larger deviations in the high pressure range
because NaCl can only provide a hydrostatic environment at low pressure range and it will induce uniaxial stress and inhomogeneous pressure distribution within the chamber of the DAC as the pressure increases~\cite{2009-tate,2016-sans}.

Through the above analysis, it can be found that the relative deviation between the results of DIA and the experimental results of Anzellini \textit{et al}.~\cite{2014-a}, Sakai \textit{et al}. ~\cite{2018-sakai} and Ono~\cite{Ono} is less than 10\% within 200 GPa.
However, the experimental results measured by Dubrovinsky \textit{et al}.~\cite{2012-du} differ significantly from other experimental and theoretical results, which leads to an overestimation of the pressure higher than $\sim$ 100 GPa.
Furthermore, there is still a large deviation between the experimental results at high pressure (see Fig.~\ref{fig3}(a-d)) even though self-consistent pressure scales are used. 

\subsection{EOS at high temperatures and high pressures}
\label{sec:2:2}
%----------------fig.4----------
\begin{figure*}
\centering\includegraphics[width=15.5 cm]{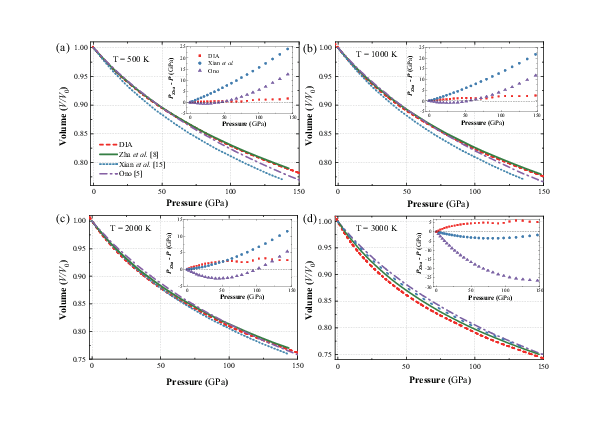}
\caption{\label{fig4} (Color Online) Isothermal dependence of relative volume $V/V_{0}$ on the pressure under 500, 1000, 2000, and 3000 K from DIA (red dashed lines), and the dataset obtained by Zha \textit{et al}.~\cite{2004-zha} (green solid lines), Xian \textit{et al}.~\cite{2022-xian} (blue dotted lines) and Ono~\cite{Ono} (purple dash-dotted lines), with $V_{0}$ the one atmospheric volume at the respective temperature. The inset shows the pressure deviation between Zha \textit{et al}. and other results.}
\end{figure*}

On the basis of the above analysis of the room-temperature situation and computational accuracy of DIA, we next turned to the EOSs of Re under higher temperature circumstances.
The isothermal EOSs were computed according to Eqs.~(\ref{eq3})-~(\ref{eq10}) and the curves at 500, 1000, 2000, and 3000 K are shown as red dashed lines in Fig.~\ref{fig4}(a)-(d) respectively, along with the corresponding \textit{P-V} data listed in Table. S1.
Since there is no direct experimental measurement of the isothermal EOSs at high temperatures, here we compared our results with other EOSs proposed by either semi-empirical method~\cite{2004-zha,Ono} or by pure theoretical calculations~\cite{2022-xian}, which are all shown in Fig.~\ref{fig4} as well.
In 2004, Zha \textit{et al}.~\cite{2004-zha} proposed a phenomenological high-temperature EOS which contains several empirical parameters that need to be fitted by experimental \textit{P-V} data points.
The pressures in this high-temperature EOS consists of two parts, the room-temperature one ($P_{\mathrm{rt}}$) and the thermal one ($P_{\mathrm{th}}$).
$P_{\mathrm{rt}}$ is described by certain empirical EOS, which, for instance, can be the third-order Birch–Murnaghan EOS determined by Duffy \textit{et al}.~\cite{1999-duff} as was used in Ref.~\cite{2004-zha}.
As to the thermal part, it is assumed that there exists a linear relationship between the $P_{\mathrm{th}}$ and temperature as $P_{\mathrm{th}}=a+bT$ for most solids in the range above the Debye temperature where the two parameters $a$ and $b$ are determined by fitting several experimental \textit{P-V} points.
The results of Ref.~\cite{2004-zha} are denoted in green solid lines in Fig.~\ref{fig4} and are in a good agreement with those by DIA.
Although the deviation between the result of DIA and Zha \textit{et al}.~\cite{2004-zha} gradually increases in the range of  0-150 GPa and 500-3000 K (see red squares in the inset of Fig.~\ref{fig4}), the largest difference is 6.1 GPa at 116 GPa and 3000 K, which is comparable to the uncertainty of common high temperature-pressure experiments ($\sim $ 5\%).

Similar to the method of Ref.~\cite{2004-zha}, Ono~\cite{Ono} also separated the pressure into $P_{\mathrm{rt}}$ and $P_{\mathrm{th}}$, where $P_{\mathrm{rt}}$ is obtained by fitting empirical EOS through experimental measurements and the $P_{\mathrm{th}}$ is directly calculated by the first-principles molecular dynamics.
As denoted as purple dash-dotted lines in Fig.~\ref{fig4}, the \textit{P-V} data obtained by Ono are basically consist with the results of Zha \textit{et al}.~\cite{2004-zha} in the low temperature range ($<$ 2000 K) while the difference from DIA is obviously larger than that of Ref.~\cite{2004-zha}.
In Ref.~\cite{2004-zha}, Zha \textit{et al}. claimed that the accuracy of the method of decomposing the pressures into $P_{\mathrm{rt}}$ and $P_{\mathrm{th}}$ critically depends on the accuracy of the isothermal EOS at the ambient temperature.
Yet, in the experiment of Ono~\cite{Ono}, the EOS measured at room temperature is in the range of 0-115 GPa and the high temperature EOS in the range of 115-428 GPa are extended results, which may substantially introduce deviations.
% At the same time, we also presented the calculation results from Ono~\cite{Ono} () and Xian \textit{et al}.~\cite{2022-xian}(blue dotted lines), and the pressure deviation between Zha \textit{et al}.~\cite{2004-zha} and the calculation results ($P_{\mathrm{Cal}}$) under the same $V/V_{0}$ are shown in the illustration by purple triangles and blue circles in the inset of Fig.~\ref{fig4}.
%----------------fig.5----------
\begin{figure}
\centering\includegraphics[width=7.5 cm]{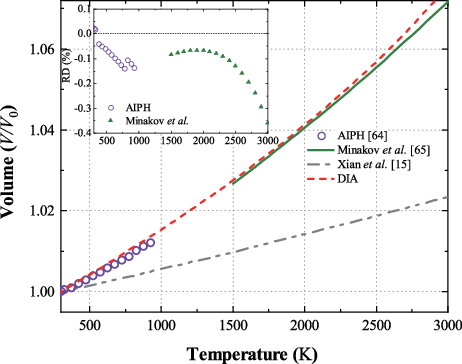}
\caption{\label{fig5}(Color Online) Isobaric EOSs of Re at one atmosphere obtained from American Institute of Physics Handbook (AIPH)\cite{1957-gray} (purple circles), DIA (red dashed line), Minakov \textit{et al}.\cite{2020-mina} (green solid line) and Xian \textit{et al}.\cite{2022-xian} (gray dash-dotted line). The inset is the relative deviation (RD) between the result of DIA and AIPH (purple open circles)/Minakov \textit{et al}. (green triangles). }
\end{figure}

In 2022, Xian \textit{et al}.\cite{2022-xian} predicted the EOS of Re up to 140 GPa and 3200 K (see blue dotted lines in Fig.~\ref{fig4}) by proposing a new thermodynamic model which consists of three parts: the two-order Rydberg-Vinet equation to describe the cold energy, the Debye model to describe the phonon contributions in free energy, and the Femi gas model to describe the free energy arising from the thermal excitation of electrons.
Although several sets of experimental data were used to constrain the parameters in the EOS formula,
the deviation between the results of Xian \textit{et al}. and other works are dramatically large even at low temperatures (see the inset of Fig.~\ref{fig4}).
Such differences become more obvious when it comes to the isobaric EOS as shown in Fig.~\ref{fig5} that the results of Ref.~\cite{2022-xian} (see gray dash-dotted line in Fig.~\ref{fig5}) exhibit large deviations from all the other experimental\cite{1957-gray} and theoretical results\cite{2020-mina}.
One possible cause of the differences, as mentioned in Ref.~\cite{2004-zha}, may be from the fact that the computational accuracy of a method involving fitting process strongly depends on the input experimental data.
Another possible reason may be resulted from the fact that the Debye model of phonon is not capable of accurately introducing the finite-temperature lattice motions as pointed out in other work~\cite{argman2015} as well as in our earlier work~\cite{2021-gong}.
On the other hand, the results by DIA (red dashed line in Fig.~\ref{fig5}) are in a good agreement with the data from American Institute of Physics Handbook (AIPH)\cite{1957-gray} (purple circles) and the theoretical results of Minakov \textit{et al}.\cite{2020-mina} (green solid line in Fig.~\ref{fig5}) based on quantum molecular dynamics (QMD), with a maximum relative deviation of 0.36\%.

\begin{figure}
\centering\includegraphics[width=7.5 cm]{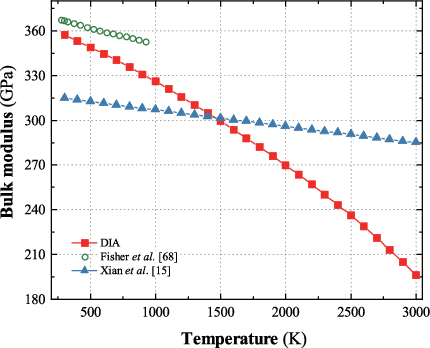}
\caption{\label{fig6}(Color Online) Bulk modulus of Re obtained by DIA (red solid squares), Xian \textit{et al}.\cite{2022-xian} (blue solid triangles), Dubrovinsky \textit{et al}.\cite{2012-du} (purple open square), Anzellini \textit{et al}.\cite{2014-a} (orange open triangle) and Fisher \textit{et al}.\cite{1967-fish} (green circles) at ambient pressure.}
\end{figure}

From the isobaric EOS obtained by DIA at one atmosphere, we calculated the bulk modulus of Re in the temperature range of 300-3000 K, as shown in Fig.~\ref{fig6} with red solid squares, which is significantly different from the theoretical result of Xian \textit{et al}.\cite{2022-xian} (blue solid triangles in Fig.~\ref{fig6}). Obviously, our calculation result is much closer to the result of the experiment performed in 1976\cite{1967-fish} (green circles in Fig.~\ref{fig6}). Recently, the experimental measurement of the bulk modulus at room temperature is about 350 GPa\cite{2012-du,2014-a} (purple open square and orange open triangle in Fig.~\ref{fig6}), which is closer to our calculation result of 357 GPa than other theoretical results. For example, Lv \textit{et al}.~\cite{2012-lv} reported a value of 389 and 376 GPa obtained by LDA and GGA respectively, while Fast \textit{et al}.~\cite{1995-fast} reported a value of 447 GPa obtained by LDA. Unfortunately, we did not find any other experimental work providing direct data on the bulk modulus of Re in the higher temperature zone and more works are certainly needed in the future.

%----------------fig.6----------

\section{Conclusion}
\label{sec:conclu}
In this work, we predicted the EOS of \textit{hcp} Re in the temperature range of 300-3000 K up to 900 GPa by solving the PF via DIA without any empirical or artificial parameters.
Since DIA is based on equilibrium statistical physics, the results should be consist with hydrostatic experiments. Unfortunately, no hydrostatic experimental results are available except for the one concerning isobaric EOS at one atmosphere, which is in good agreement with our calculated result.
Compared with the results of quasi-hydrostatic compression experiments at room temperature, our calculated results present relative deviations of less than 10\% below 200 GPa.
Furthermore, the semi-empirical high-temperature EOSs up to 150 GPa generally agree well with our results.
These facts indicate that the EOSs of Re calculated by DIA can be a reliable reference, and more future experimental researches on Re at high temperatures and high pressures are needed to verify our predictions.
\section{Acknowledgement}
BYN acknowledges the financial supports from the Key Laboratory of Thorium Energy of CAS.
% \bibliography{ref}
%merlin.mbs apsrev4-1.bst 2010-07-25 4.21a (PWD, AO, DPC) hacked
%Control: key (0)
%Control: author (72) initials jnrlst
%Control: editor formatted (1) identically to author
%Control: production of article title (-1) disabled
%Control: page (0) single
%Control: year (1) truncated
%Control: production of eprint (0) enabled
%
\end{document}